\documentstyle [12pt]{article}
\newcommand{\be}{\begin{equation}}
\newcommand{\ee}{\end{equation}}
\newcommand{\bees}{\begin{eqnarray}}
\newcommand{\ees}{\end{eqnarray}}
\newcommand{\ra}{\rightarrow}

\begin{document}

\begin{titlepage}
\begin{flushright}
 IFUP-TH 8/96\\
  February  1996\\
  gr-qc/9603010
\end{flushright}

\vspace{5mm}

\begin{center}

{\Large\bf  Symmetry breaking aspects of the effective \\

\vspace{3mm}

Lagrangian for quantum black holes}

\vspace{12mm}
{\large Alessandra Buonanno$^{a,b}$, Stefano Foffa$^{a}$,
Michele Maggiore$^{b,a}$\\}

\vspace{2mm}

and {\large Carlo Ungarelli$^{a,b}$}

\vspace{3mm}

(a) Dipartimento di Fisica dell'Universit\`{a},\\
piazza Torricelli 2, I-56100 Pisa, Italy.\\
(b) Istituto Nazionale di Fisica Nucleare, sezione di Pisa, Italy.
\end{center}

\vspace{4mm}

\begin{quote}
\hspace*{5mm} {\bf Abstract.} The physical excitations entering the
effective Lagrangian for quantum black holes are related to
a Goldstone boson which is present in the Rindler limit
and  is due to the spontaneous breaking of the translation symmetry
of the underlying Minkowski space.
This physical interpretation, which 
closely parallels similar well-known
 results for the effective stringlike description of flux tubes in
QCD, gives a physical insight into the problem of
describing the quantum degrees of freedom of black holes.
It also suggests that the recently suggested concept of 'black hole
complementarity' emerges at the effective Lagrangian level rather
than at the fundamental level.

\end{quote}
\end{titlepage}
\clearpage

{\bf 1. Introduction}

\vspace{3mm}

The attempts to give a description of black holes  consistent
with the laws of quantum mechanics face well known problems. A
possible approach~\cite{tH,Sus} assumes that at the quantum level,
from the point of view of an external, static observer, the 
quantum degrees of
freedom of a black hole are located on the horizon (see
also~\cite{memb}).
In this approach, because of the blue-shift factor in quantities like the
Hawking temperature, a static observer sufficiently close to the horizon 
is in a region of super-Planckian energies, where 
unknown physics comes into play.  To describe quantitatively the
horizon dynamics one can therefore resort to an effective Lagrangian
approach~\cite{MM1,BGMPU}. The most general
effective action turns out to be of the form
\be\label{Seff}
S_{\rm eff}  =  -{\cal T}\int d^3\xi\sqrt{-h}\, \left[ 1+
C_0 K+
C_{\scriptscriptstyle I}R+
C_{\scriptscriptstyle II}K^2+ 
C_{\scriptscriptstyle III}K_{ij}K^{ij}+\ldots \right]\, .
\ee
The basic variables appearing in the action are 
the fields $\zeta^{\mu}(\xi )$,  $\xi^i=(\tau ,\sigma_1 ,\sigma_2)$
which define the position of the quantum, fluctuating horizon and
describe a 2+1 dimensional timelike hypersurface (the
world-volume)
parametrized by $(\tau ,\sigma_1 ,\sigma_2)$ and
embedded in 3+1 dimensional spacetime with background metric
$g_{\mu\nu}$. From these one constructs the induced metric
$h_{ij}$ and the extrinsic curvature $K_{ij}$ which appear in
eq.~(\ref{Seff}), where $h=\det h_{ij}$ and $K=K_i^i$; $R$ is the
scalar curvature of the world-volume. The coefficients of the various
operators are phenomenological constants which can in principle be
derived if one knows the underlying fundamental theory.

In a semiclassical expansion, the nature of the degrees of freedom
entering  eq.~(\ref{Seff}) is more transparent. In flat space one
can write the generic fluctuation over a classical solution 
$\bar{\zeta}^{\mu}(\xi )$ as
\be\label{phi}
\zeta^{\mu}(\xi ) -\bar{\zeta}^{\mu}(\xi )=b^i(\xi
)\partial_i\bar{\zeta}^{\mu} +\phi (\xi)\bar{n}^{\mu}(\xi )\, ,
\ee
where $\partial_i =\partial /\partial\xi^i$ and 
$\bar{n}^{\mu}(\xi )$ is the normal of the classical solution
$\bar{\zeta}^{\mu}(\xi )$ in the point of the world-volume labelled by
$\xi$. One  observes~\cite{GV}
that the fields $b^i(\xi )$ represent fluctuations along the surface 
and are 'pure gauge' since they can be
reabsorbed with a reparametrization. The only physical quantum
fluctuations 
are  perpendicular to the surface, and are parametrized by a
single scalar field $\phi (\xi )$ living in the world-volume. 
The definition can be generalized to curved space, where the physical
fluctuations are written as
\be
\zeta^{\mu}(\xi )=
\zeta^{\mu}_{\rm geod}(\phi ,\bar{\zeta}, \bar{n}) \, ,
\ee
where, at each point $\xi$ on the 
world-volume, $\zeta^{\mu}_{\rm geod}$ gives the geodesic
  parametrized by an affine parameter
$\phi$, which at $\phi =0$ goes through the point 
$\bar{\zeta}^{\mu}$, with a tangent at $\phi =0$ equal to
$\bar{n}^{\mu}$.

Expanding at quadratic order in $\phi$, in generic curved space,
the action becomes~\cite{Carter,BGMPU} 
\be\label{lin}
S_{\rm eff}=S_{\rm cl}-\frac{{\cal T}}{2}\int d^3\xi \sqrt{-\bar{h}}\left[
\partial_i\phi\partial^i\phi - (\bar{K}_i^j\bar{K}_j^i+
\bar{{\cal R}}_{\mu\nu}\bar{n}^{\mu}\bar{n}^{\nu})\phi^2\right]\, ,
\ee
where the overbar denotes the value at $\bar{\zeta}$,
$S_{\rm cl}=-{\cal T}\int d^3\xi (-\bar{h})^{1/2}$,
 indices are raised and lowered with $\bar{h}_{ij}$, and
$\bar{{\cal R}}_{\mu\nu}$ is the Ricci tensor of the embedding space
evaluated at  $\bar{\zeta}$.

For a planar membrane in Rindler space the term $\sim\phi^2$ vanishes and we
are left with a massless scalar field living in the 2+1 dimensional
world-volume. The appearance of a massless scalar particle in an effective
Lagrangian leads naturally to suspect that we have to
do with a Goldstone boson. Indeed, this turns out to be correct, and
it is in fact well known in the context of the string description of
chromoelectric flux tubes in {\em QCD}~\cite{Lus}. To our knowledge,
however, this has not been properly appreciated in the studies on
quantum fluctuations of domain walls or membranes, and since it gives 
interesting hints on the problem of quantization of black holes, 
we find useful to
discuss it in the present context.

\vspace{3mm}

{\bf 2. Toy Model}

\vspace{2mm}

To understand why a Goldstone boson appears, let us see in an
explicit example 
how an effective membrane theory  emerges from a fundamental
theory. As the fundamental theory we take
\be\label{fund}
S=-\frac{1}{2}\int d^4x\,
\left[\partial_{\mu}\Phi \partial^{\mu}\Phi 
+g\left(\Phi^2-\frac{m^2}{g}\right)^2\right]
\ee
in flat space, $g_{\mu\nu}=(-,+,+,+)$.
The theory has different  sectors depending
on the boundary conditions that we impose. In the sector defined by
$\Phi(z\ra +\infty )=m/\sqrt{g}, \Phi(z\ra -\infty )=-m/\sqrt{g}$ 
we have a manifold of 
ground states given by the domain wall solutions,
\be
\Phi_{\rm cl} =\frac{m}{\sqrt{g}}\tanh m(z-z_0)\, ,
\ee
labelled by a parameter $z_0$, which  is the collective coordinate
corresponding to translation invariance in the direction transverse to
the domain wall.
If we select a particular member of this manifold of ground states in
order to perform a semiclassical expansion, we are  breaking
spontaneously the translation invariance along $z$ and we expect to find a
corresponding Goldstone boson. Expanding the field $\Phi (x)=
\Phi_{\rm cl}(x)+\eta (x)$, the action for the fluctuations is~\cite{Lus}
\be\label{c1}
S= S_0 -\frac{1}{2}\int d^4x\, \eta (x)\Delta^{\Phi}\eta (x)
\ee
\be\label{c2}
\Delta^{\Phi}=-\partial_{\mu}\partial^{\mu} +V(z)
\ee
\be\label{c3}
V(z)=4m^2-\frac{6m^2}{\cosh^2m(z-z_0)}
\ee
Therefore the eigenfunctions have the form
$\eta (x) =e^{i(k_0t+k_1x+k_2y)}\psi (z)$
where $\psi (z)$ satisfies a one-dimensional Schroedinger equation
\be\label{Sch}
\left( -\partial_z^2+V(z)\right)\psi (z)=\epsilon\psi (z)
\ee
and the eigenvalues are $-k_0^2+k_1^2+k_2^2+\epsilon$. 
The Schroedinger equation has two bound states~\cite{Lus}
\begin{eqnarray}\label{spectrum}
\psi_0(z)&=&\frac{m^2}{\sqrt{g}}\,
\frac{1}{\cosh^2m(z-z_0)}\, ,\hspace{10mm}\epsilon =0\\
\psi_1(z)&=&\frac{m^2}{\sqrt{g}}\,
\frac{\sinh mz}{\cosh^2m(z-z_0)},\hspace{10mm}\epsilon =3m^2\, ,
\end{eqnarray}
and a continuum of modes $\psi_{k_3}$
which starts at $4m^2$, $\epsilon =4m^2+k_3^2$. The normalization of
the modes has been chosen for later convenience.

The crucial point is the existence of a mass gap separating the mode
$\psi_0$ from the rest of the spectrum. This means that, if we are
interested in low energy physics, $E^2<3m^2$, we can integrate out all
modes except the mode $\psi_0$. Expanding $\eta $ in normal modes
and using the notation $\xi =(t,x,y)$
\be
\eta (\xi ,z)=c_0(\xi )\psi_0(z)+c_1(\xi )\psi_1(z)
+\int dk\, c_k(\xi )\psi_k(z)
\ee
the integration in $dz$ in the action, eq.~(\ref{c1}), can be
performed explicitly and
the action $S$ becomes a functional of the fields
$c_0(\xi ),c_1(\xi ),c_k(\xi )$ living in the world-volume.
The effective action for the mode $c_0$ is obtained integrating over
all massive modes,
\be
\exp{(iS_{\rm eff}[c_0])}=N\int Dc_1\prod_kDc_k\, e^{iS}\, ,
\ee
where $N$ is a normalization factor.
At quadratic order in $\eta$ the normal modes decouple and
\be\label{c0}
S_{\rm eff}= -\frac{{\cal T}}{2}\int d^3\xi\,
\,\partial_i c_0 \partial^i c_0\, ,
\ee
where
\be
{\cal T}=\int_{-\infty}^{\infty} dz\, \psi_0^2(z)=\frac{4m^3}{3g}\, .
\ee
Expanding at higher order in $\eta$ we obtain a
coupling between the various modes, which generate higher dimension
operators in the effective lagrangian for $c_0$. These terms, however,
are irrelevant in the low energy limit.

The fact that a mass term in eq.~(\ref{c0}) is absent can be
understood noting that
 $\psi_0(z)=\partial_z\Phi_{\rm cl}$ and then
\be\label{gold}
\Phi_{\rm cl}(t,x,y,z)+c_0(t,x,y )\psi_0(z)=\Phi_{\rm cl}(t,x,y,z+
c_0(t,x,y))+O(c_0^2)\, .
\ee
Therefore infinitesimal  rigid translations
in the $z$ direction are realized on the 
world-volume field $c_0$ as
$c_0(\xi )\ra c_0(\xi )+$ const., and this symmetry of the embedding space 
forbids the presence
of a mass term in the effective action for $c_0$. 
We see that $c_0(\xi )$ 
is a Goldstone boson which lives in the world-volume 
of the domain wall and is
associated to the spontaneous breaking of translation symmetry.
Note that this field propagates only along the membrane, since it has 
$k_3=0$, and the associated mode $\psi_0(z)$ is localized around the
membrane, and it determines its thickness through the parameter $m$.

To understand the relation between $c_0(\xi )$
and the field $\phi (\xi)$ defined in eq.~(\ref{phi}) let us see how,
in the same toy model defined by eq.~(\ref{fund}), the effective
lagrangian~(\ref{Seff}) can be explicitly derived.
The technique was discussed in ref.~\cite{For} and a
systematic  evaluation of higher order terms has been
presented in ref.~\cite{CG}. The idea is to separate explicitly the
dependence on the transverse direction of the quantities which appear
in the action~(\ref{fund}), so that we can integrate it out. 
The first step is to choose an appropriate coordinate system
suited to the  domain wall that we are
considering, which is taken to be a small fluctuation over a planar solution.
Thus, in our flat space example, rather then using
cartesian coordinates $(t,x,y,z)$ we use three coordinates 
$\xi^i =(\tau,\sigma_1,\sigma_2)$
which parametrize the world-volume and, as a  coordinate
in the transverse direction, we use 
the affine parameter $\lambda$ which parametrizes
the geodesic which pass through the point of the domain wall labelled
by $\xi$ and is orthogonal to the domain wall. At least in a
neighbourhood of the domain wall this coordinate system is well
defined, and this is all we need when considering small fluctuations
around a planar wall.  In this coordinate system the normal 
is $n^{\mu}=(0,0,0,1)$ even if the wall is non planar. Next 
one introduces the tensors $h_{\mu\nu}=g_{\mu\nu}-n_{\mu}n_{\nu}$ and
$K_{\mu\nu}=n_{\mu ;\nu}$ where the
semicolon denotes the covariant derivative.
They satisfy relations which can be obtained as follows. Computing the Lie 
derivative of $h_{\mu\nu}$ in this coordinate system one finds
\be\label{b1}
\frac{\partial h_{\mu\nu}}{\partial\lambda}=2K_{\mu\nu}\, ,
\ee
which we take as our first fundamental equation.  Furthermore, in flat
space, the Riemann tensor is zero. We write explicitly the equation
for the component $3\mu 3\nu$ in this coordinate system, where
$x^{\mu}=(\tau ,\sigma^1,\sigma^2,\lambda )$
\be
0={\cal R}^{3}_{\,\mu 3\nu}=\Gamma^{3}_{\mu\nu
,3} -\Gamma^{3}_{\mu 3 ,\nu}+
\Gamma^{3}_{\rho 3}\Gamma^{\rho}_{\mu\nu}-
\Gamma^{3}_{\rho\nu}\Gamma^{\rho}_{\mu 3}\, ,
\ee
where $\Gamma$ is the Christoffel symbol and the comma is the ordinary
derivative. Observing that in this coordinate system
$K_{\mu\nu}=n_{\mu ,\nu}-\Gamma^{\rho}_{\mu\nu}n_{\rho}=-
\Gamma^{3}_{\mu\nu}$ and that similarly $K_{\mu}^{\nu}=
\Gamma^{\nu}_{\mu 3}$ and $\Gamma^{3}_{\mu 3}=0$, we get
\be\label{b2}
\frac{\partial K_{\mu\nu}}{\partial\lambda}=K_{\mu}^{\rho}
K_{\rho\nu}\, ,
\ee
which is the second fundamental equation. The equation
of motion for $\Phi$, written separating explicitly the transverse and
longitudinal parts, reads
\be\label{b3}
\frac{\partial^2\Phi}{\partial\lambda^2}+K
\frac{\partial\Phi}{\partial\lambda}+\nabla_i\nabla^i\Phi-
2g\Phi (\Phi^2-\frac{m^2}{g})=0\, .
\ee
Eqs.~(\ref{b1},\ref{b2},\ref{b3}) are now expanded in powers of the small
parameter $\varepsilon =l/L$ 
where $l=1/m$ is the thickness of the wall and $L$ is
the typical lengthscale over which the world-volume bends ($L=\infty$
for an exactly planar wall). The condition that the wall be a small
perturbation over the planar solution is implemented requiring
\be
l\frac{\partial\Phi}{\partial\lambda}=O(1)\, ,\hspace{10mm}
l\frac{\partial\Phi}{\partial\xi^i}=O(\varepsilon )\, .
\ee
Introducing the dimensionless quantities $u=\lambda/l ,v^i=\xi^i/L,
\Psi =(\sqrt{g}/m)\Phi$ and $k_{\mu\nu}=LK_{\mu\nu}$ one obtains~\cite{CG}
\bees
\frac{\partial h_{\mu\nu}}{\partial u}&=&2\varepsilon k_{\mu\nu}
\label{a1}\\
\frac{\partial k_{\mu\nu}}{\partial u}&=&\varepsilon k_{\mu\rho}
k_{\nu}^{\rho}\label{a2}
\ees
\be\label{a3}
\frac{\partial^2\Psi}{\partial u^2}-2\Psi(\Psi^2-1)+
\varepsilon k \frac{\partial\Psi}{\partial u}+
\varepsilon^2 {\cal D}_i{\cal D}^i\Psi =0\, ,
\ee
where ${\cal D}_i$ is the covariant derivative on the world-volume,
with respect to the rescaled
variables $v^i$. Each of the quantities appearing
in eqs.(\ref{a1}-\ref{a3}) can now be expanded in $\varepsilon$,
e.g. 
\be\label{psi}
\Psi =\Psi_0 +\varepsilon\Psi_1 +\frac{\varepsilon^2}{2}\Psi_2+\ldots
\ee
and the equations can be solved order by order in $\varepsilon$. This
allows to determine explicitly the dependence on $u=\lambda/l$. Then, writing
the original action in the form
\be
S=\int d^3\xi d\lambda\sqrt{-g}\, {\cal L}\, ,
\ee
where $\sqrt{-g}$ is the Jacobian for the transformation from the
cartesian coordinates $(t,x,y,z)$  to $(\xi^i,\lambda )$, the integral
over $\lambda$ can be performed explicitly and the result~\cite{CG} is
eq.~(\ref{Seff}). Of course the computation gives also an explicit
expression for the constants $C_0,C_I,$ etc. in eq.~(1). In particular,
one finds $C_0=0$ because  it is given by the integral of an odd
function of $u$ from $-\infty$ to $+\infty$ (note however that
$C_0\neq 0$ if we work in a finite volume $-L_1\leq z\leq L_2$), and 
$C_{\scriptscriptstyle III}$ can be set to zero
since in flat space  the operator $K_{ij}K^{ij}$ is not independent
of $R$ and $K^2$ because of the Gauss-Codacci identity.
The other coefficients turn out to be\footnote{We have found here a
numerical discrepancy in ${\cal T},C_{\scriptscriptstyle II}$
with the result quoted in ref.~\cite{CG}. For comparison,
the quantities $\lambda ,\eta$ 
in ref.~\cite{CG} are denoted here $g/2$ and $m/\sqrt{g}$
respectively.}
\bees
{\cal T}&=&\frac{4m^3}{3g}\\
C_{\scriptscriptstyle I}&=&\frac{1}{m^2}\,\frac{\pi^2-6}{24}\\
C_{\scriptscriptstyle II}&=&-\frac{1}{m^2}\,\frac{5}{48}\, .
\ees
Eq.~(\ref{Seff}) can now be expanded  around a 
classical solution $\bar{\zeta}^{\mu}$ and,
if we consider only the leading term $\sim\sqrt{-h}$
 the result~\cite{Carter,BGMPU}
is given by eq.~(\ref{lin}).
For a planar domain wall in flat space the mass term in
eq.~(\ref{lin}) vanishes.

\vspace{2mm}

{\bf 3. Implications}

\vspace{3mm}

The conclusion from this exercise is as follows. In the toy model
defined by eq.~(\ref{fund}) all computations can be performed
explicitly. We can explicitly derive the effective Lagrangian,
eq.~(\ref{Seff}), including the numerical value of the
phenomenological constants, 
we can introduce a field $\phi$, eq.~(\ref{phi}),
 parametrizing physical quantum fluctuations, and we can derive the
action which governs its dynamics, which is just the action of
a massless scalar field living
in a world-volume with  $\bar{h}_{\rm ij}=(-,+,+)$.
 On the other hand,
we can  solve the spectrum of the fluctuations
of $\Phi$, eqs.~(\ref{c2}),  and we see
explicitly what is the reason which allows to use an effective
Lagrangian approach: it is the existence of a Goldstone boson,
separated by a mass gap from the rest of the spectrum. This allows to
quantify explicitly what  does it mean low energy in the effective
Lagrangian approach: it means $E^2<3m^2$. In the range $3m^2<E^2<4 m^2$
we must take into account also the mode $c_1(\xi )$ 
and above $4m^2$ we have
the continuum. The comparison of eqs.~(\ref{lin}) 
(with $\bar{h}_{\rm ij}=(-,+,+)$, $\bar{K}_i^j\bar{K}^i_j=0$ and
$\bar{{\cal R}}_{\mu\nu}=0$)
and (\ref{c0}) shows
that $\phi(\xi )=c_0(\xi )+O(c_0^2)$, i.e.
in the limit of small fluctuations
$\phi(\xi )$ is nothing but the Goldstone mode $c_0(\xi )$. The
identification does not extend to  finite fluctuations. This can be
seen  observing that for a planar membrane in flat space the
translation symmetry $z\ra z+a$ is implemented on $\phi$ as
$\phi\ra\phi+a$ exactly, as we read from eq.~(\ref{phi}) 
setting $n^{\mu}=(0,0,0,1)$. Instead $c_0$ transforms as $c_0 \rightarrow c_0+a$ only
for infinitesimal values of $a$, as we see from
eq.~(\ref{gold}). To obtain a representation of a finite translation,
all modes $c_k$ must be taken as a basis, and they transform
non-linearly between themselves.

Let us see what  can we learn from the above discussion in the case
of the effective Lagrangian for quantum black holes.
 Of course in this case we do not know the
fundamental theory from which  eq.~(\ref{Seff}) should emerge.
In the approach of refs.~\cite{MM1,BGMPU} one therefore {\em postulates}
 that, for a
static observer outside the horizon, the variables $\zeta^{\mu}(\xi )$
which describe the position of the quantum, fluctuating horizon are
the relevant degrees of freedom at low energies; the
action~(\ref{Seff}) then follows from symmetry considerations. In the
following we limit ourselves to the leading term in eq.~(\ref{Seff}),
\be\label{memb}
S_{\rm memb}=  -{\cal T}\int d^3\xi\sqrt{-h}\, .
\ee
Let us consider first the Rindler metric, which is the limit of the
Schwarzschild metric for large black hole mass at a fixed distance
from the horizon, and is the metric appropriate for an observer with
constant acceleration $g$ in Minkowski space. 
We denote Minkowski coordinates by $(T,x,y,Z)$ and we define 
Rindler coordinates $t,z$ from
$T=z\sinh gt\, , Z=z\cosh gt$; this mapping only covers 
the wedge $|Z|\ge |T|, Z\ge 0$ (see Fig.1). 
The Minkowski metric becomes in Rindler coordinates $g_{\mu\nu}=
(-g^2z^2,1,1,1)$ and in these coordinates the equation of motion 
of the action~(\ref{memb}) has a  solution~\cite{MM1} of the form
\be\label{sol}
\bar{\zeta}^{\mu}(\xi )=(\tau ,\sigma_1,\sigma_2,
\frac{z_0}{\cosh g\tau})\, .
\ee
while, using the Minkowski coordinates, it takes the form 
\be\label{sol1}
\bar{\zeta}^{\mu}(\xi )=(z_0\tanh g\tau ,\sigma_1,\sigma_2,
z_0)\, .
\ee
If we now expand the
action around this solution  we get eq.~(\ref{lin}) with
the  term $\sim\phi^2$ equal to zero~\cite{BGMPU}. In the variables
$\xi^i=(\tilde{\tau}=z_0\tanh g\tau ,\sigma_1,\sigma_2)$ we have
$\bar{h}_{ij}=(-1,1,1)$ and therefore 
the equation of motion is the massless Klein-Gordon equation
in a flat space with boundaries in the temporal direction, since
$|\tilde{\tau}|\leq z_0$.  The action governing the field $\phi$ is
therefore invariant under the field transformation $\phi (\xi)\ra\phi
(\xi )+a$. From the previous discussion, we are lead to ask whether
this transformation can be interpreted as a symmetry operation in the
embedding space where the (unknown) fundamental theory lives.
The interesting point is that, if we limit ourselves to the Rindler
wedge defined above, there is no such a symmetry. In fact,
an infinitesimal  transformation $\phi\ra\phi +a$ generates a
translation along the $Z$ axis in Minkowski space since, 
$\zeta^{\mu} (\xi)=\bar{\zeta}^{\mu}+\phi\,\bar{n}^{\mu}$
and the normal  $\bar{n}^{\mu}$ to the classical solution~(\ref{sol1})
points along the $Z$ axis, but such transformation is not a  symmetry
for the Rindler wedge.

However, if we consider the full Minkowski space, rather than the
Rindler wedge, then the transformation
$\phi\ra\phi  +a$ is associated to the symmetry of
translation along the $Z$ axis in the embedding space. 
Thus, in analogy with the toy model, eq.~(\ref{fund}), the field
$\phi$ can be related to a Goldstone boson if the underlying
theory lives in the full Minkowski space, i.e. the maximum analytical 
extension of the Rindler wedge.

This simple observation gives the following suggestion. In
the spirit of black hole complementarity~\cite{Sus} one tries to
describe a quantum black hole, from the point of view of an external,
static observer, without making any reference to what happens inside
the horizon. The degrees of freedom of the black hole are taken to
live in a small region  near the horizon, the so called 
'stretched horizon'~\cite{Sus,MM2}.  The above discussion, however,
suggests that if we look for a fundamental theory which in the low
energy limit  (i.e. at sub-Planckian energies)
reproduces the effective action for quantum black holes, we
cannot limit ourselves to the region outside the horizon. The
fundamental theory, in the Schwarzschild case, must live in the
maximum analytic extension of the Schwarzschild space (or
in Minkowski space if we work in the Rindler limit).

Such a theory should be defined without reference to any particular
observer. 
It is only when we try to derive an effective theory from this
fundamental theory that a dependence on the observer appears. A static
observer outside the horizon will obtain his effective action
integrating over the fundamental variables in
the region from where he cannot receive signals.
A free falling observer, instead, can receive signals from any region
and therefore he cannot define an effective action.

It is this procedure which introduces
an observer dependence in the low energy theory. It appears
therefore that the 'tension' between the point of views of a static
observer and a free falling observer, which has been termed
'black hole complementarity'~\cite{Sus} is something which emerges at
an effective, rather than at a fundamental level.

As a final remark we observe that the identification of $\phi$ 
with a Goldstone boson has been done in the Rindler limit; 
in the Schwarzschild  case, instead, the field 
$\phi$ is not massless,
although the effective mass term vanishes in the limit of large black hole
mass~\cite{BGMPU}. 
However, the idea that the
fundamental theory should live in the maximal analytical extension
and that black hole complementarity only emerges at an effective
level should be of rather general validity and should carry over from
the Rindler limit to the Schwarzschild case.

\begin{figure}
\begin{picture}(300,300)(0,0)
\put(-110,-220){\includegraphics{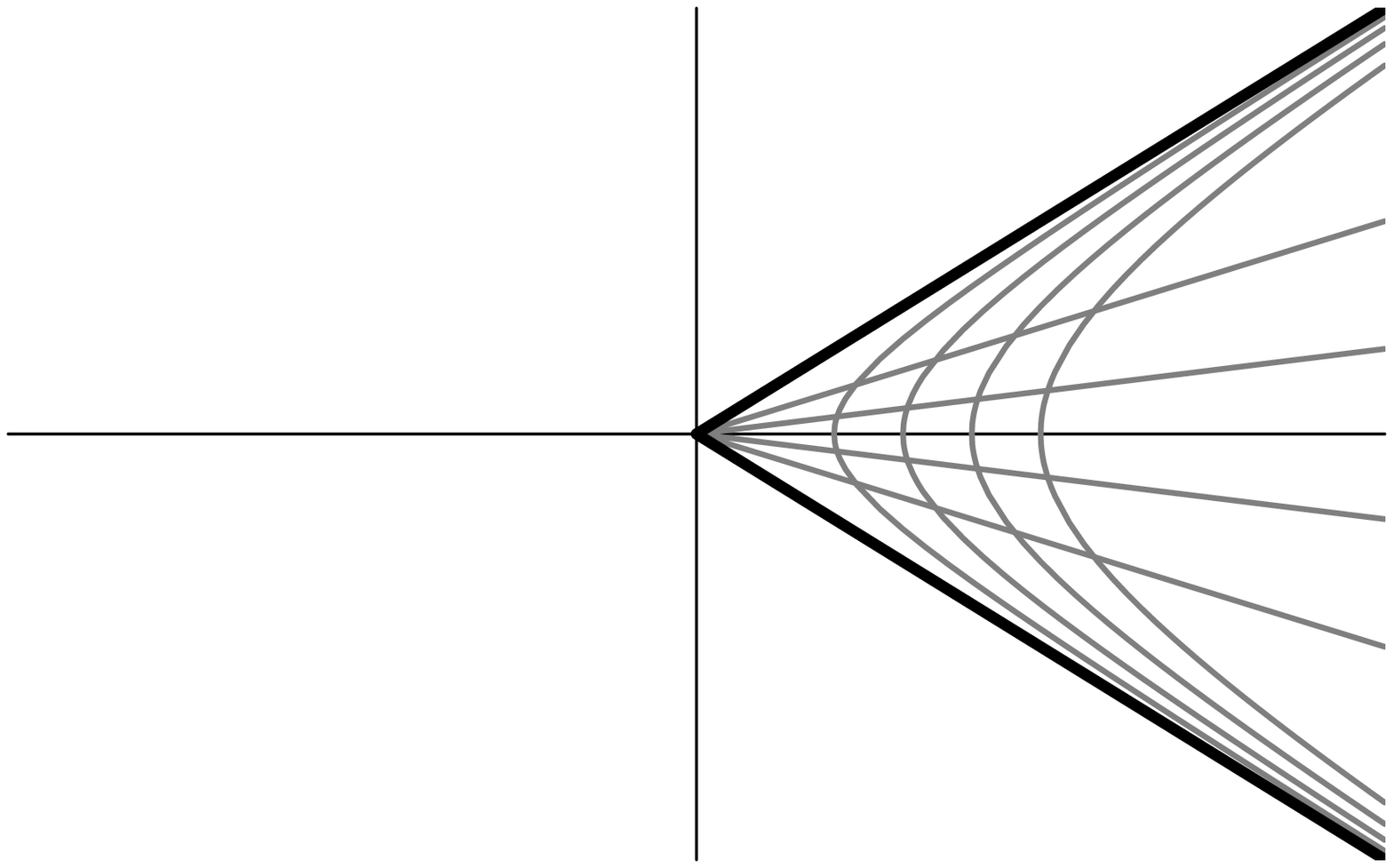}}
\end{picture}
\put(125,165){$Z$}
\put(-110,330){$T$}
\put(-150,-100){Fig.1: Rindler wedge}
\end{figure}
\end{document}